\documentclass[a4paper]{article}
 
\usepackage{amssymb}
\usepackage{amsmath}
\usepackage{latexsym}

\usepackage{shadow}

\usepackage{hyperref}

\usepackage[toc, page]{appendix}
\usepackage{appendix}
\usepackage{url}

\def\dof {degree of freedom}
\def\dofs {degrees of freedom}
\def\guill {\textquotedblleft ~}
\def\guillr {\textquotedblright}
\def\spt {space-time}
\def\gr {general relativity}
\def\vf  {vector-field}
\def\eg {\emph{e.g.}}
\def\fb {fiber bundle}
\def\CC {{\bf C}}
\def\calD {{\cal D}}
\def\blit {\emph}
\def\R{{\rm I\!R}}
\def\dmanif {differentiable  manifold}
\def\coord {coordinate}
\def\Vol {\mbox{Vol}}
\def\eqbydef {~\overset{\text{def}}{=}~}
\def\d {{\rm d}}
\def\Sect {\mbox{Sect}}
\def\YY {{\bf Y}}
\def\calC {{\cal C}}
\def\jet {{\cal J}}
\def\wrt  {w.r.t.~}
\def\Hodge {{\star}}
\def\hodge {{\star}}
\def\T{{\rm T}}
\def\QQ {{\bf Q}}
\def\Image {\mbox{Im}}
\def\calY {{\cal Y}}
\def\calH {{\cal H}}
\def\calP {{\cal P}}
\def\calS {{\cal S}}
\def\calT {{\cal T}}
\def\D {{\rm D}}
\def\etc {{\it etc.}}
\def\ie {\emph{i.e.}}
\def\Poinc{Poincar{\'e}}
\def\calL {{\cal L}}
\def\undemi {\tfrac{1}{2}~}
\def\munu { {\mu \nu } }
\def\VV {{\bf V}}
\def \ELequ  {Euler-Lagrange equation~}

\newcommand{\arxiv} [1] { \url{http://arxiv.org/abs/#1} }  
\newcommand{\arXiv} [1] { \url{http://arxiv.org/abs/#1} }  

\newcommand{\dEL}[2] {\frac{\delta^{ \tiny  EL}  #1}{\delta #2}}

\newcommand{\see}[1]{(see~ \ref{#1})}
\newcommand{\brak}[1]{\langle #1  \rangle}
\newcommand{\pder}[2]{\frac{\partial#1}{\partial#2}}

\newcommand\innerp[1] { #1  \mathbin\lrcorner} 
\newcommand{\eql}[2]
{ \begin{equation} \label{#1} 
 #2
\end{equation}}
\newcommand{\eqn}[1]{equ.(\ref{#1})}
\newcommand{\eqbox}[2]
{ \begin{equation} \label{#2}   
   \boxed{ 
 #1   }
\end{equation}}

\title{Historical Hamiltonian  Dynamics: symplectic and covariant} 
\author{M. Lachi{\`e}ze-Rey, \\
 APC - Astroparticule et Cosmologie (UMR 7164) \\
 Universit\'e Paris 7 Denis Diderot
 10, rue Alice Domon et L\'eonie Duquet \\ F-75205 Paris Cedex 13, France\\
mlr@apc.univ-paris7.fr
}
\begin{document}
\maketitle
 
{\bf Abstract}

This paper presents a   \guill historical   \guillr    ~formalism for dynamical systems, in its Hamiltonian version (Lagrangian version was presented in a previous paper).
It is universal, in the sense that it applies equally  well to time dynamics and to field theories on \spt.   
It is based on the    notion of (Hamiltonian)  \emph{histories}, which are sections of the (extended) phase space bundle. It is developed in the space of sections,  in contradistinction with the usual formalism which works in the bundle manifold. 

In field theories, the  formalism remains   covariant    and   does not require a spitting of \spt.  
It  considers \emph{\spt} ~exactly in  the same manner than \emph{time} in usual dynamics, 
both being  particular cases of the \emph{evolution domain}. It   applies without modification  when   the histories   (the fields)   are forms rather than scalar    functions, like in     electromagnetism or in  tetrad      \gr.
 
We develop a differential calculus in the infinite dimensional space of histories.  It  admits    a (generalized)  symplectic form  which does not break the covariance. We develop  a covariant symplectic formalism, with generalizations of  usual notions  like current conservation, Hamiltonian \vf s,   evolution \vf, brackets, ... The usual  multisymplectic approach    derives form it, as well as  the symplectic form introduced by Crnkovic and   Witten 
in the space of  \emph{solutions}.  
       
\section{Introduction}\label{intro} 
 
Our historical   Hamiltonian formalism is based on the notion of \emph{history}.    According to \cite{Sorkin},    histories \guill  furnish the raw material from which reality is constructed\guillr.

This follows our previous work  (\cite{paperI}, hereafter  paper I)   which presents a \emph{ Lagrangian} formalism on the same basis (see an outline in Appendix \ref{outline}).
An history (or kinematical history) is a \emph{possible} evolution of a dynamical system, also called a \emph{configuration} \cite{MarsdenPekarskyShkollerWest}.  
An history which obeys the dynamical equations becomes  a \emph{physical} evolution, or particular solution.

Our approach  applies  equally well to usual  time dynamics (tD) and to  covariant field theories (FT), and allows further generalizations.
These different contexts  (tD and FT) only differ by their \emph{evolution domain} (see below):  the time line in tD; the  \spt ~in FT. But all expressions or  equations are identical in both cases. 
Thus \spt ~in FTs appears on the same footing than time in tD, with the only difference that it is 4 dimensional  rather than mono-dimensional. 
In the case of FTs, our formalism    remains entirely covariant and  does not require  any splitting of \spt.  

In addition,  it  applies without modification to       the case where the    fields are not \emph{functions}, but \emph{forms} (\eg, on \spt). This applies to  electromagnetism or to  \gr ~in the tetrad formalism.

In paper I, we have presented its    Lagrangian  version. Here we present the     Hamiltonian one. 
The most   important result is the existence of a canonical  (generalized)  symplectic form which remains entirely covariant. In  time dynamics, it is equivalent   to the usual symplectic form. In field theories, it remains  covariant and we show that     the multisymplectic formalism may be seen  as     derived from it.  
To mention some general ideas underlying this approach, 
\begin{itemize}
  \item  dynamics is  defined, not  versus time  but versus an \emph{evolution domain}. It  reduces to  the time line in  tD (a particular case);  to    \spt ~for FT's. But it  is treated exactly in the same manner in both cases.

  \item An history may be   a function on the evolution domain (like a scalar field on \spt) but also, more generally, 
   a  differential  form on it; in any case  a section of a particular  \fb. 
  \item A \emph{particular solution} is an history which is  an orbit  of a [Hamiltonian] flow in the corresponding  bundle. Such flows have   the   dimension    of the evolution domain. They    may be called  \emph{general solutions}. 
  \item Our calculus does not hold in configuration space, or phase  space, but 
in the space of histories which has   infinite dimension It is inspired by  \emph{diffeology} \cite{PIZ} considerations. It may be seen as   generalization of both  \cite{Witten}  and   the multisymplectic formalism, and as a synthesis between  them.
 
  \item In  the  space of (Hamiltonian)  histories, we define  a  canonical and covariant (generalized)  symplectic form; equivalent to  the usual symplectic form in tD; 
  giving raise to  both  the multisymplectic form and   the symplectic currents \cite{Witten} in  FTs.
  \item
  Covariant field theories (in \spt)  and time dynamics appear  as  two particular cases  of this formalism. 
\end{itemize}
 
 In some sense, our formalism appears as a   generalized synthesis between   the    multisymplectic geometry  (see, \eg, \cite{Helein}), the \guill covariant phase space \guillr  approaches (see, \eg, \cite{Helein}),  the canonical approach and the geometry of the space of solutions. It remains entirely covariant.

The section \ref{General}  introduces 
 the notion of history  (\ref{history}). It  defines their prolongations  
 to \emph{velocity-histories} and  \emph{ Hamiltonian histories}, involved in the Lagrangian and Hamiltonian formulations of  dynamics, and introduces the  \emph{phase space bundle} 
(\ref{Hamiltonian Histories}).  It also explicits our differential calculus in the space of Hamiltonian Histories (\ref{calcul}).
Section \ref{Dynamics} expresses the Hamiltonian Dynamics in its historical formulation. It 
introduces the generalized symplectic form and the evolution H-vector field   (\ref{symplectic}).
It 
derives the universal historical evolution equations (\ref{evolution}) and explicits the dynamical solution (\ref{solution}). 
Section  \ref{illustrations} gives illustrations, applying the general formalism to 
{time Dynamics}
(\ref{time Dynamics}),
and to scalar field theories, where our formalism is compared to the 
multisymplectic one (\ref{Multisymplectic}).
Section \ref{Conservation and symmetries} considers  conservation (\ref{On shell conservation}) and symmetries  (\ref{symmetries})
It  discusses the notions of [generalized]  Poisson brackets and observables  (\ref{Poissonobservables}).
The last sections apply  to  electromagnetism  
(\ref{electromagnetism}) and to first order \gr ~(\ref{GR}).
An outline of 
the historical Lagrangian formalism presented in paper I is given in Appendix \ref{outline}.
  

\section{General framework}\label{General}

Our framework applies equally well to (non relativistic) time dynamics and to relativistic (covariant) field theories. It is formulated in terms of \emph{histories}, that we define below. Shortly, an history is associated to each \dof ~of the dynamical system. We treat only the case of an unique \dof; the generalization to multicomponent-systems is straightforward and is treated in   illustrations below. The case of a form-field rather than a scalar field, like the electromagnetic potential in Maxwell theory, is treated as a single  \dof, 
to which correspond an unique (although not scalar-valued) history (a  1--history); or the tetrad \gr, where
 the cotetrads forms $e^I$ and   and the spin connection   forms $\omega^{IJ}  $ are also 1--histories.
We treat first the scalar field case and then extend  to the form field. 

\subsection{Histories}\label{history}

A    dynamical  system is characterized by its
configuration bundle $\CC \to \calD$.\\~Here, $\calD$ is the \blit{domain} of the theory. In  usual time dynamics (tD), this is the time line $  \R_t$, or an interval of it. In  relativistic  field theories (FT), 
this is  \spt. We treat both cases  equally, and more generally, $\calD$ is a n-dimensional \dmanif,  possibly with a given metric.  We label $\calD$ with  arbitrary coordinates $x^\mu$  (the  unique \coord ~$t=x^0$ for the timeline in tD), which     disappear in  our final results which are     covariant
\footnote{ We define the non covariant forms $\Vol\eqbydef \d ^n x$, $\Vol_\mu\eqbydef \innerp{\partial_\mu }\Vol$ 
and $\Vol_{\mu \alpha}\eqbydef \innerp{\partial_\alpha ,\partial_\mu }\Vol$ in $\calD$, as usual.}.  They  generate  adapted [local] \coord s in the various \fb s we will consider. 
Our philosophy is  to treat $\calD$ as some kind of   \guill n-dimensional timeline~" \wrt which the evolution is expressed.  

An element of the fiber is a possible value of the dynamical variable.
Most physical  systems admit   many \dofs ~(or components). We treat the case of an  unique component. 
The generalization to   composite fields is straightforward as it will appear in the examples below.
Thus for  the particle in space, an history   corresponds to   each    \coord ~as  $C:~t\to q^i(t)$; 
for a \emph{scalar} field in \spt, $C: ~(x^\mu) \to C(x^\mu)$ generally written $ \varphi (x^\mu)$; 
for a composite   field, one history for each component $\phi ^A$.

An history (or  {field-history}, or configuration), that we always write  $C$,  is a section 
\footnote{ For mathematical conditions imposed on them, see, \eg, 
\cite{Barbero}.} of the configuration bundle  $\CC$:
 a function on $\calD$ for the particle or  for the scalar field; but, more generally, a differential form on $\calD$ like in electromagnetism or in tetrad \gr ~(see below).
Thus, the   \emph{space of histories}   
 $\calC=\Sect(\CC)$, or possibly a subset of it.
The histories which obey the dynamical equations are the particular solutions.

\subsection{The  Phase Space Bundle and Hamiltonian Histories}\label{Hamiltonian Histories}

Given an history  $C$, the corresponding   \blit{velocity-history}   is its  first jet extension (or prolongation),  the pair $C_V\eqbydef jC\eqbydef (C, \d C)$ (with $\d$  the 
 exterior derivative in~$\calD$) or $(C,C_\mu)$   in components. This is a section of the first jet bundle $\jet \CC$.
In paper I, we have developed the Lagrangian historical formalism in this jet bundle \see{outline}.  
\footnote{ An interesting different point of view \cite{HigherprequantumgeometrySchreiber} considers a field configuration as a section of the infinite jet bundle $\jet ^\infty \CC$.}

Its  affine dual 
$\jet ^* \CC\to \CC$. Its bundle manifold,  the phase space
\footnote{ Different  authors use  various appellations   for  this bundle or for its associated  manifold: the  covariant phase space bundle,  the doubly extended phase space 
\cite{PauflerRomer},    the extended dual bundle \cite{SarletaWaeyaert}, 
the extended multimomentum bundle \cite{RomanRoy}, the \blit{  De Donder-Weyl multisymplectic manifold}
...}, admits the  adapted [Darboux]    \coord s 
\footnote{ For time dynamiccs, replace  $x^\mu$ by $t$, $ \phi$  by $q$, $p^\mu$ by $p$.}
  $x^\mu, \phi,p^\mu,\pi$. 
They act by duality \cite{MarsdenPekarskyShkollerWest}   as $$\brak{(x^\mu, \phi,p^\mu,\pi), (x^\mu, \phi,v_\mu )}=(p^\mu~v_{\mu}+\pi)~\Vol.$$
We see the polymomenta  $p^\mu$  as   the  dual components of     the (n-1)--form over~$\calD$,  $p\eqbydef p^\mu~\Vol _\mu$, that we call the polymomentum
\footnote{ Equivalently, the $p^\mu$ are  the components of the dual  polymomentum $\Hodge p=p^\mu~Ê\d x ^\mu$ (sum over indices).}.

The (extended)  phase space bundle is the bundle $\YY=\jet ^* \calC\to\calD$.
A   section is a map 
$$Y=(X^\mu,C,P,\Pi):~x^\mu\to X^\mu( x^\mu)= x^\mu,C( x^\mu),P( x^\mu),\Pi( x^\mu),$$ 
that we call an  Hamiltonian history (hereafter H-history)
\footnote{ It is known that  $\YY$ may also be seen as the bundle $\bigwedge ^n_2 \T^* \QQ$ of $n$-forms over~$ \QQ$ which annihilates two arbitrary vertical \vf s, see, \eg, \cite{Helein}, \cite{Lawson}.
In this case, $p^\mu$ and $\pi$  appear  as the coefficients in the expansion of  such an $n$-form. 

 There is a canonical projection 
 which projects it  out  to the \emph{linear}Ê~dual \cite{PauflerRomer} $ \widetilde{\YY}$,
 forming the  line bundle  \cite{SarletaWaeyaert} 
 $$\rho:~\YY \to 
 \widetilde{\YY}:~(x^ \mu, \varphi, p^\mu,\pi)\to (x^ \mu, \varphi, p^\mu).$$
 Interestingly \cite{SarletaWaeyaert,CendraCapriotti,ForgerSalles},  the (scalar) Hamiltonian may be seen as a section $\widetilde{h}$ of that bundle, which defines the function $H$ on $\widetilde{\YY}$ through 
 $$\widetilde{h}(x^ \mu, \varphi, p^\mu)
=~(x^ \mu, \varphi, p^\mu,H~\left(x^ \mu, \varphi, p^\mu)\right) .$$
It is equivalent   to work in $\YY $ or in  $\widetilde{\YY}$.    
  Both are polysymplectic. For the relation between both approaches, see also  \cite{CendraCapriotti,SarletaWaeyaert,ForgerSalles}.}. 
The  components are expressed in the table \ref{componentsT}, where    $\Omega^k_D=\Omega^k(\calD)$ is the space of k-forms on $\calD$. 
We call $P$     the \blit{historical momentum}.    The trivial maps    $X^\mu$, 
     defined for convenience,  will appear as the conjugate variables to the $\Pi_\mu$.

\begin{table}[h]
\caption{The components of an hamiltonian history}
\begin{center}
\begin{tabular}{|c|c|c|}
\hline
&&\\
$C$ & $\calD\to \Omega _D^0$ & $(x^\mu)\to C(x^\mu)$      \\
&&\\
\hline
&&\\
$P=P^\nu~\Vol_\nu$ & $\calD\to \Omega _D^{n-1}$ & $(x^\mu)\to P (x^\mu)$  \\
&&\\
$P^\nu$ & $\calD\to \Omega _D^0$ & $(x^\mu)\to P^\nu(x^\mu)$      \\
&&\\
\hline
&&\\
$\Pi_\nu$ & $\calD\to \Omega _D^{n-1}$ & $(x^\mu)\to \Pi_\nu(x^\mu)$     \\
&&\\
$\Pi=\Pi_\nu~Ê\d x^\nu$ & $\calD\to \Omega _D^{n }$ & $(x^\mu)\to \Pi (x^\mu)$     \\
&&\\
\hline
&&\\
$X^\nu~Ê\d x^\nu$ & $\calD\to \Omega _D^{0 }$ & $(x^\mu)\to X^\nu(x^\mu)=x^\nu$  \\
&&\\
\hline
\end{tabular}
\end{center}
\label{componentsT}
\end{table}%

Any H-history $Y$  defines a n-dimensional hypersurface in the phase space, which is simply its image $\Image (Y)$.
And $Y$  is a diffeomorphism $\calD\to \Image (Y)=Y(\calD)$.
When the history is  a solution, $\Image (Y)$ is an orbit of the evolution flux (see below).

We will work in the space of  Hamiltonian histories rather than in the phase space bundle. 
We first  define differential  calculus in it.

\subsection{Extension to form-fields}

Our formalism applies equally well in  the case where a field-history $C$ is  a r-form, rather than a function (0-form), on $\calD$.
We treat explicitly   the case  $r=1$. This applies to  electromagnetism, where $C$ corresponds to  the Maxwell potential~$A$;  or to  general relativity in tetrad formalism, where  histories correspond to  the cotetrad fields $e^I$ and to the connection forms $\omega^{IJ}$, see below. 
We do not consider   separately the  components of a form-field,  but we treat  it globally  as an history    $C$ as in the table \ref{componentsMM}.

The scalar case corresponds to $r=0$. 
When $r>1$, the treatment is similar, with indices replaced by multi-indices (see paper I, and appendix \ref{multiindex}).
The table \ref{componentsMM} presents the components of an Hamiltonian history in the   case  $r=1$.
{\bf In all formula, juxtaposition implies wedge product in $\calD$}.
 We calculate now in  the infinite dimensional space  $\calY$ of Hamilton--histories.

\begin{table}[h]
\caption{The components of an hamiltonian history}
\begin{center}
\begin{tabular}{|c|c|c|}
\hline
&&\\
$C=C_\alpha   \d x^\alpha$ & $\calD\to \Omega _D^r$ & $(x^\mu)\to C(x^\mu)=C_\alpha (x^\mu) ~Ê\d x^\alpha$    \\
&&\\
\hline
&&\\
$P=P^{\mu \alpha}  ~ Ê\Vol_{\mu \alpha}  $ & $\calD\to \Omega _D^{n-1-r}$ & $(x^\mu)\to P (x^\mu)=
P^{\mu \alpha}  (x^\mu) ~  Ê\Vol_{\mu \alpha} $ \\
&&\\
\hline
\end{tabular}
\end{center}
\label{componentsMM}
\end{table}%

   \subsection{Differential Calculus  with  Hamiltonian histories }\label{calcul}
 
An Hamiltonian history (H-history)  $Y$ is a  section  of the bundle $\YY$.  We call~$\calY$ the infinite  dimensional space of   H-histories, and we construct differential calculus on it
\footnote{  Note that similar approaches   (\cite{Zuckerman}, \cite{Deligne})   consider  elements of $\Omega (\Sect(\Omega_D \times \calD)$.}.
We represent such a section (a H-history, a \guill point " of   $\calY$) as 
$$Y=(X,C,P,\Pi) =(Y^A),$$ where we treat  the $Y^A= X,C,P,\Pi$    (with $A=1,2,3,4$)   like  four  coordinates in~$\calY$\footnote{ $X$  holds for the four $X^\mu$;  $C$ holds for the infinite set of values $C(x)$ (or $C_\mu (x)$ if $r \ne 0$). Our notation allows us to manage this infinite set  like one unique \coord; similarly with~$P$. }.


We   generalize the notions of functions, \vf s, differential forms... to  H-maps, H-\vf s, H-forms.
This appears necessary to define a correct calculus. 
A     H--map is an   application      $$F:~\calY\to \Omega (M):~ Y=(Y^A) \to F(Y) = F(Y^A).$$
When $ F(Y)\in \Omega^R (M)$, we call $F$ a [0;R]-map.
The    Hamiltonian functional~$\calH$ will appear as  a particular   [0,n]--map. 
We write  $\calC (\calY)=\Omega ^0(\calY)$ the space of H-maps.

Hereafter, juxtaposition of H-maps will mean  their wedge product on $\calD$, always implicit. 
This gives to  $\calC (\calY)$  an algebra structure.
Also, the differential calculus on $\calD$ is easily lifted to $\calY$ through the formula 
$$(\d F) (Y)= \d (F(Y)).$$
We call  occasionally $\d$ the \emph{horizontal} derivative, but we do not consider it as part of the  proper differential calculus on $\calY$. We    introduce below a genuine  external derivative $\D$ in $\calY$, different from $\d $ and commuting with it. This  is analog to the double complex structure  introduced by  \cite{Deligne}.


  \subsubsection{ Derivations are \vf s}

We first define     derivations of H-maps  \wrt  their   arguments $Y^A$,  under the form of      basic   partial  derivative operators $\partial _{ A}=\pder{}{Y^A}$    acting on $\calY$. 
This is accomplished   through the variation formula  (wedge product in $\calD$ assumed)
\eql{variationformula}{\delta  F
=\pder{F}{ Y ^A} ~\delta Y ^A    
=\pder{F}{ X} ~\delta X
+   \pder{C}{ X} ~\delta C
+\pder{F}{ P} ~\delta P
+\pder{F}{ \Pi} ~\delta \Pi}
$$
=\pder{F}{ X^\mu} ~\delta X^\mu
+   \pder{C}{ X} ~\delta C
+\pder{F}{ P} ~\delta P
+\pder{F}{ \Pi_\mu} ~\delta \Pi _\mu
, $$
corresponding to
 the general variation of a H-history
$$\delta Y=(\delta X,\delta C, \delta P, \delta \Pi) =(\delta Y^A) .$$


We call the   operators $\partial _{ A}$ the basic   H-\vf s in $\calY$. 
The general H-\vf ~on $\calY$  is $V=V^A~ \partial_A  $, 
whose   components $V^A\in \calC (\calY)$ are arbitrary H-maps. 
It acts  on an   H-map $F$,   as  
 $V(F)=V^A ~Ê \pder{F}{Y^A}  $ (wedge product in $\calD$  still implicit) \footnote{ This requires some conditions on the grade of $F$ that we do not detail here.}.

 \subsubsection{H-forms}

We   define [differential] H-forms in $\calY$ through duality.
First    the    \blit{basis one-forms} 
$\D  Y^A$ -- which mean  the collection 
$  \D X^\mu,\D C, \D P, \D \Pi_\mu $ -- through their actions on an arbitrary H-\vf,
$$\brak{\D  Y^A,  V}=V^A.$$
The \emph{general  one-H-form} expands as 
$$\alpha=\alpha_A  ~\D Y ^A, $$
whose  components  $\alpha_A\in \calC(\calY)  $  
are arbitrary H-maps (sum over repeated indices  is always assumed).    We have  $$\brak{\alpha,V}=\alpha_{A}~V^A;$$
 and the exterior derivative of a H--map $F$
 $$\D F=\pder{F}{Y ^A}~\D Y ^A.   $$ This  is just  an other way to write  \eqn{variationformula}, after realizing that a variation of a H-history  is simply the action of a H-\vf ~$\delta=\delta^A~Ê\partial _A$ on it, namely
 $$\delta Y^A=\delta (Y^A)=\brak{\D Y^A, \delta}=\delta^A $$  (this requires   $\delta ^A$ 
to be of  the same  grade  than $Y^A$).
   When $F$ is a [0,R]-map, we call $\D F$ 
 a [1,R]-H-form; [0,R]-maps are [0,R]-H-forms.  

The wedge product of H-forms,  $\wedge$ (not to be confused with the wedge product on $\calD$ which is always implicit), is defined as antisymmetrized tensor product, as usual. 
It generates [2,R]-H-forms, \etc ~The external derivative $\D$ also applies to [k,R]-forms and generates [k+1,R]-forms.
Thus we have the rules
expressed in table \ref{expla}.
 \begin{table}[h] 
\caption{Differentials of  H-forms}
\begin{center}
\begin{tabular}{|c|}
\hline \label{expla}
  ~\\
  ( {[k;R]-form})~$ \wedge$~ ( {[k';R']-form}) = {[k+k';R+R']-form}    \\
  ~\\
$\d$ ( {[k;R]-form}) = {[k;R+1]-form} ;  \\
  ~\\
$\D$ ( {[k;R]-form}) = {[k+1;R ]-form} . \\
  ~\\
\hline
\end{tabular}
\end{center}
\end{table}%
 Contraction of  H-\vf s with H-form is as usual.
 
 We have for instance 
 $$\D \calH= 
 \pder{\calH }{X^\mu}~\D X^\mu
+ \pder{\calH }{C}~\D Y ^C
 +\pder{\calH }{P}~\D Y ^P
+ \pder{\calH }{\Pi_\mu}~\D Y ^{\Pi_\mu} $$
$$ =
  \pder{\calH_0}{C}~\D Y ^C
  +\pder{\calH_0}{P}~\D Y ^P
+\d x^\mu ~\D Y ^{\Pi_\mu}
 $$ where we used \eqn{HamiltonianHform} in the last term.
 
 These formulas also apply equally well  in the case where the histories are not scalar, \ie,  [0,r]-histories rather than [0,0]-histories. 
We give
in appendix \ref{Details} their explicit development   for
  one-form valued 
 histories, \ie,  [0,1]-histories rather than [0,0]-histories.
 They generalize easily to the general case of 
 [0,r]-histories.
    We give in table \ref{gradeHH} the grades of the different H-maps and H-forms involved (scalar case corresponds to $r=0$).
        
\begin{table}[h]
\caption{The grades  of   the H-maps and H-forms}
\begin{center}
\begin{tabular}{|c|c|c|c|c|c|c|} \hline
&&&&&\\
H-form &   $\calH$& $\pder{\calH}{C}$   &$\pder{\calH}{P}$&$\D \calH$&$\Omega$\\
&&&&&\\
\hline
&&&&&\\
grade  &[0;n ]&[0,n-r]& [0;r+1]  &[1;n]&[2;n-1]\\
&&&&&\\
\hline\end{tabular}
\end{center}
\label{gradeHH}
\end{table}%

\section{Dynamics and evolution}\label{Dynamics}

\subsection{The symplectic H-form}\label{symplectic}

The space of histories $\calY$ admits  the canonical   [1;n-1]-Hform 
$$\Theta \eqbydef P~\D C+\Pi_\mu ~\D X^\mu
=   P^\mu~ \D C ~\Vol_\mu +\pi  ~\D X^\mu ~\Vol_\mu  ,$$
that we call the  \Poinc-Cartan H-form
\cite{MarsdenPekarskyShkollerWest}.
      Its   [vertical] exterior derivative $$\Omega\eqbydef \D \Theta = \D P\wedge \D C
+ \D \Pi_\mu \wedge  ~\D X^\mu ,$$
 is a  closed and non degenerate [2;n-1]--form on $\calY$, that we call  the \blit{symplectic H--form}.
For a 0-history,   
 $ \Omega =(\D P^{ \mu} \wedge \D C)~\Vol_\mu $; for a 1-history,   
 $ \Omega =(\D P^{\alpha\mu} \wedge \D C_\alpha)~\Vol_\mu  $. For  $r>1$, the same formulas hold, with  indices  replaced by 
multi--indices.   
 We will see below that it allows us to construct a genuine  (scalar-valued)  symplectic form \emph{in the   space of solutions}, which identifies to that   of \cite{Witten}.

Under some conditions, a H--map $F$ admits a \blit{symplectic gradient} $\nabla_\Omega F$, a  H-\vf ~defined through 
$$\innerp{\nabla_\Omega F} \Omega=\D F.$$ 

\subsection{Evolution }
\label{evolution}

The Dynamics is described by the  \blit{historical  Hamiltonian}\footnote{ see appendix \ref{outline} for obtaining it as a result of a Legendre transform.} 
\eql{HamiltonianHform}{\calH=h~\Vol=H_0(C,P)+  \Pi_\mu~\d x^\mu.} 

This is a  [0;n]-H-map
$\calH:~\calY\to \Omega^n_D$.

 The \blit{evolution \vf} is 
 defined  as its    {symplectic gradient} $Z=\nabla_\Omega   \calH$:
$$\innerp{Z} \Omega = \D \calH.$$ 
We emphasize that there is no analog in the multisymplectic formalism. It expands   as  $Z=Z^A~Ê\partial_A$,
and the   equation above gives its  components through  $Z^A~Ê\Omega _{AB}Ê=\pder{\calH}{Y^B}$, with explicit solution
\eql{components}{
Z^ PÊ=-\pder{\calH}{C},~~~~
Z^ CÊ=  \pder{\calH}{P},~~~~~~~~
Z^{X^\mu}=\pder{\calH}{\Pi_\mu}=\d x^\mu,~~~~~~
Z^{\Pi_\mu}=-\pder{\calH}{X^\mu}=0 }
(note the difference between $x^\mu$ and $X^\mu$),
where $Z^P$ and $Z^C$ are [0;n-r]- Êand [0,r+1]--Hmaps respectively. 
 This {evolution \vf}   acts as a derivation operator on any H--map $F$,  giving  the  H--map $$Z(F)=
Z^A ~Ê\pder{F}{Y^A},$$
with the components given in \eqn{components}; 
in particular the derivatives of the \guill  \coord s ", $Z(Y^A)=Z^A$. 
In particular
$Z(X^\mu)=\d x^\mu$, $Z(\Pi^\mu)=0$.

 \subsection{The dynamical  solution}\label{solution}

An H--history $Y=(Y^A)$ is a real motion (solution) when the evolution \vf ~is tangent to it. This  means 
 $\d Y^A= Z(Y^A)=Z^A $, \ie,  
using  (\ref{components}) 
\eqbox{\d C =\pder{\calH}{P};~~~\d P=-\pder{\calH}{C},~~~~~\d X^\mu =\d x^\mu,~~~~~\d \Pi_\mu =0. }{universal}
The two last are identities.
We recall that   $\d$ is  the (horizontal) 
exterior derivative  in~$\calD$, not be confused with   exterior derivative  $\D$ in $\calY$.

This     \guill historical "   version of the 
Hamilton--De Donder--Weyl equations applies to tD as well to FT. We show below that it leads to the usual dynamical  equations. It includes the case where 
the field  is a form rather than a map (\eg, electromagnetism or \gr), as we show in applications below.  For a multi--component history (field) it holds for each component.

It is easy to check that the previous equations  insures  stationarity of the action $\int _{\calD}~\calL$, with the Lagrangian H--map (see paper I) $$\calL =P~Ê\d C -\calH.$$
Namely,
using the commutativity between $\d$ and $\D$, 
$$\D \calL
 = \D (P~Ê\d C -\calH)   = \D  P~Ê\d C -    P~\D Ê\d C -\D \calH  =$$
$$= \D  P~Ê\d C - \epsilon  (\d (    P~\D Ê  C)- \d      P~\D Ê  C)  - (\pder{\calH}{C} ~\D C +\pder{\calH}{P} ~\D P) .$$
Inserting the motion equations above, this reduces to $\D \calL = -  \d (    P~\D Ê  c)$,  an exact form in $\calD$ which   gives zero contribution to the integral, QED.

 
\section{Illustrations}\label{illustrations}
\subsection{     Application to    time Dynamics}\label{time Dynamics}

In usual dynamics, $\calD$ is the time line, $\Vol =\d t$,   
$\Pi =\pi ~Ê\d t$ and   $\Pi _\mu =\Pi _t =\pi$.
Then
$$\calH = \undemi \Hodge P ~ÊP + U(C)~\d t + \pi  ~Ê\d t =    h~Ê\d t,$$  
with $h= \undemi  P ~ÊP ~Ê  + U(C)~  + \pi~ $ the usual Hamiltonian \emph{function}; $C=q$ and $P=p$ are  zero-forms ($r=0$).
Then  $\pder{\calH}{C} =\pder{h}{C} ~Ê\d t= U'(C) ~Ê\d t $ 
and 
$\pder{\calH}{P}=\pder{h}{P}~Ê\d t  =P ~Ê\d t $  
are both  [0;1]-Hmaps. 
$\Omega = \D P\wedge \D C + \D \Pi \wedge \D T $ is a [2,0]-form (a genuine scalar valued  symplectic form).

Then, 
  (\ref{universal}) immediately gives  the usual Hamilton  equations  (we reintroduce the familiar notations):
$$ \d C =\dot{C}  ~Ê\d t =\pder{\calH}{P}=\pder{h}{P}~Ê\d t \implies
  \dot{C}   =  \dot{q}   =  \pder{h }{p}.$$

  $$\d P=\dot{P} ~Ê\d t  =- \pder{\calH}{c}=- \pder{h }{c}~Ê\d t \implies
  \dot{p}   =- \pder{h }{c};$$
with 
\eql{evolx}{\d T=\pder{\calH}{\Pi_t}=\d t ;~~~~ \d \pi =\pder{\calH}{T}=0.}

\subsection{Scalar field;  Link with Multisymplectic}\label{Multisymplectic}

For classical field theories, 
 $\calD=M$ is \spt ~($n=4)$. A scalar  ($r=0$) field   $C$ is usually written  $\varphi$. Then   $P= P^\mu~Ê\Vol _\mu$ is a [0,3]-Hmap, with \emph{dual} components $P^\mu$, $\calH=h_0~\Vol +\Pi_ \mu ~\d x^\mu $  is a [0,4]-Hmap. 
We have $$ \d C = C_{,\mu}~Ê\d x^\mu ,~
P = P^{\mu}  ~Ê\Vol_\mu ,~
\d P = P^{\mu}_{~,\alpha}~\d x^\alpha~Ê\Vol_\mu = P^{\mu}_{~,\mu}~ \Vol .$$

  Then   
$\pder{\calH}{P}    =\pder{h}{P^\mu}~\d x^\mu$ is a [0,1]-Hmap; 
$\pder{\calH}{C} =  \pder{h}{C}~\Vol$ is a  [0;4]-Hmap. The symplectic  [2,3]-Hform $$\Omega = \D P\wedge \D C +   \D \Pi_\mu \wedge \D X^\mu= \D P^\mu  \wedge \D C ~Ê\Vol _\mu+ \D \Pi_\mu \wedge \D X^\mu,$$
  with $\Vol _\mu  $   a 3-form on  \spt ~$\calD=M$ (not on $\calY$).
Then, \eqn{universal} implies  the   usual Hamilton equations 
$$C_{,\mu}  =  \pder{h}{P^\mu};~~~~~ 
       (P^\alpha)_{,\alpha}    =   - \pder{h}{C}   .$$ 


Assuming the standard Hamiltonian for scalar field theories, 
 $$H=\undemi \Hodge P ~P+U(C)~\Vol+\undemi \Hodge \Pi _\mu ~\d x^\mu=
 ( \undemi   P^\mu ~P_\mu~ +U(C)~\Vol+\undemi \pi^2)~\Vol,$$ we obtain
 $$\d C=C_{,\mu}~Ê\d x^\mu = \Hodge P=P^\mu ~Ê\d x^\mu \Rightarrow
C_{,\mu}=P^\mu ;$$
 $$\d P =-\pder{\calH}{C}=-U'(C)~Ê\Vol\implies
 \d P^\mu~\Vol_\mu=-U'(C)~Ê\d x^\mu~\Vol_\mu \implies
    P^\mu_{,\mu}= C_{,\mu\mu} =-U'(C).$$ 

\subsubsection{Link with Multisymplectic}

The multisymplectic form appears as an emanation of our symplectic H-form, as  the  5-form in the phase space bundle 
manifold $\YY$  (not on $S^Y$),    

$$\Omega _M =  {~\overset{\tiny{-}}\d}  p^\mu~Ê {~\overset{\tiny{-}}\wedge } ~VOL _\mu  
{~\overset{\tiny{-}} \wedge }~
 {~\overset{\tiny{-}}\d}
 \varphi+ {~\overset{\tiny{-}}\d} \pi   {~\overset{\tiny{-}} \wedge }~
    VOL, $$ where  all forms, exterior derivative 
${~\overset{\tiny{-}}\d}$ and wedge product ${~\overset{\tiny{-}}\wedge }$ are  in the bundle manifold $\YY$,   $VOL   \eqbydef \epsilon_{\mu\alpha\beta\gamma}  ~\overset{\tiny{-}}\d x^\mu  {~\overset{\tiny{-}}\wedge }   ~~\overset{\tiny{-}}\d x^\alpha  {~\overset{\tiny{-}}\wedge } ~~ \overset{\tiny{-}}\d x^\beta   {~\overset{\tiny{-}}\wedge } ~~ \overset{\tiny{-}} \d  x^\gamma$ and \\
$VOL _\mu \eqbydef \epsilon_{\mu\alpha\beta\gamma}  ~\overset{\tiny{-}}\d x^\alpha  {~\overset{\tiny{-}}\wedge } ~ \overset{\tiny{-}} \d x^\beta   {~\overset{\tiny{-}}\wedge }  \overset{\tiny{-}} \d  x^\gamma$.

 \subsubsection{  Application to      r-histories} 
Exactly the same formalism applies when fields are forms rather than scalar functions, with indices replaced by multi-indices \see{multiindex}:
$$  c =  c_{\underline{\alpha} }~Ê\d x^{\underline{\alpha}} ,~
 \d c =  c_{\underline{\alpha},\mu}~Ê\d x^{\underline{\alpha}\mu} ;$$ 
$$P = P^{\underline{\alpha}\mu}  ~Ê\Vol_{\underline{\alpha}\mu} ,~
\d P = P^{\underline{\alpha}\mu}_{~~~,\beta}  ~Ê\Vol_{\underline{\alpha}\mu}~\d ^\beta ,~
 \calH = h~\Vol,$$
 $$  \pder{\calH}{P}=\pder{h}{P^{\underline{\alpha}\mu}}~\d  ^ {\underline{\alpha}\mu},~\pder{\calH}{c}=\pder{h}{c^{\underline{\alpha}}}~\Vol_{\underline{\alpha}},$$ 
  giving the   Hamilton equations 

 $$c_{\underline{\alpha},\mu}  =  \pder{h}{P^{\underline{\alpha} \mu}};~~~~~ 
        (P^{\underline{\alpha} \mu})_{,\mu}    =   - \pder{h}{c _   {\underline{\alpha}}}   ,$$ where all multi-indexes are antisymmetrized.

\section{Conservation and symmetries}\label{Conservation and symmetries}
\subsection{On shell conservation}\label{On shell conservation}

Interestingly, \eqn{universal} implies, \emph{on shell},  
$$\D \calH =\pder{\calH}{c}~Ê\D c +
 \pder{\calH}{P} ~Ê\D P  \simeq \d c ~Ê\D P-\d P~\D c  $$ 
$$ \implies \D \D \calH =0= \D \d c ~Ê\D P-\D \d P~D c= \d  {\Omega}  $$   after derivation: the generalized symplectic form is conserved on shell. This is the covariant version   of the  on shell conservation of the symplectic current in the multisymplectic formalism.

Since the value of $ {\Omega}$ is a $(n-1)$-form on $\calD$, it can be integrated along  a $1$-codimensional  hypersurface of $\calD$. This 
 provides a canonical  scalar-valued symplectic form on the space of solutions since    the on--shell conservation of  $ {\Omega}$   implies that this symplectic  form does not depend on the choice of the hypersurface (assumed Cauchy for FTs). Thus, this provides  a canonical (scalar valued) symplectic form  on the space of histories, which identifies with that introduced by \cite{Witten}, so that our result may be seen as a generalization of their work and  its  link with the multi--symplectic formalism.

  \subsection{Symmetries}\label{symmetries}

We recall that a solution is a  H-history $Y$ verifying $Z(Y) =\d Y$ or, in \coord s, $Z^A=\d Y^A$.
Any \vf ~$\delta$ (of convenient grade) defines a variation $\delta (Y)$ of that history. One may check immediately that   the variation of a solution remains a solution, \ie, that 
$$Z(Y) =\d Y\implies  Z(\delta(Y)) =\d \delta(Y).$$

\begin{itemize}
  \item 
A symmetry is a   Hamiltonian  \vf ~$\delta$ that preserves $H$:  
  $$0=\delta (H)=\innerp{\delta} \D H =\innerp{\delta} (\innerp{Z} \D H) 
  =-\innerp{Z} (\innerp{\delta} \omega) =\omega(\delta,Z).$$
In \coord s, this implies $\delta^A~\pder{H}{Y^A}=0$.

     \item 
Being   Hamiltonian, $\delta$  is a  symplectic gradient:
$$\innerp{\delta} \omega =\D U.$$
Then 
$$ \delta (H)=-\innerp{Z} (\D U)=-Z(U)=0: $$ the quantity $U$ is conserved  on shell.
\end{itemize}
  
\subsection{Generalized Poisson bracket and observables}
\label{Poissonobservables}

The main result here is the introduction of the historical symplectic H--form~$\Omega$. Is it possible to define a Poisson-like bracket from it ?
The formula above suggests that the canonical \guill variables " are the forms  $C$ and $P$ and that  the  bracket of two Hmaps could be defined  as
$$\{f,g\}=\pder{f}{C}~\pder{g}{P} - \pder{g}{C}~\pder{f}{P}=\innerp{X_f}  \D g,$$ 
 involving the multisyplectic gradient $X_f$ such that $\innerp{X_f}  {\Omega}=\D f$.

\begin{table}[h]
\caption{The types of the  Hmaps and Hforms involved}
\begin{center}
\begin{tabular}{|c|c|c|c|c|c|c|c|}
\hline
c&P&$f, \D f$ & $g, \D g$&  $\Omega$ & $X_f$& $\{f,g\}$\\
&&&&&&\\
$[0;r]$&[0;n-r-1]&[0;R],[1;R]  &  [0;S],[1;S] &  [2,n-1] &  [-1;R+1-n]  &[0,S+R+1-n]\\
\hline
\end{tabular}
\end{center}
\label{TYPES}
\end{table}%

We give in  table \ref{TYPES} the
grades of the various quantities involved. The grade  [-1;R+1-n]  for the \vf ~indicates that the inner product with a [1;K]--Hform gives a     [0;R+1-n+K]--Hmap. 
This definition requires that the quantities involved are well defined and we restrict the validity of our bracket  to such  cases. This occurs when $f$ and $g$ have both degrees greater or equal to those of~$c$ and~$P$, namely $r$ and $n-r-1$; or, alternatively, when $f$ or $g$ does not depend on the \guill canonical  variables ". To illustrate, we have 
$$\{P,c\}=1;$$
$$\{\calH,c\}= \pder{\calH}{P}=\d c;$$
$$\{\calH,P\}=-  \pder{\calH}{c}=\d P.$$
These formulas  validate  the definition of our bracket. 
It  is a generalization of  that proposed  by \cite{Kanatchikov}.

It is defined for Hmaps, whose values are forms, rather than  scalar   functions. However, an observable is generally considered as scalar-valued, not form-valued. But any form provides a scalar  by integration over a submanifold  of adapted dimension. 
Thus, it  seems a convenient point of view   to consider generalized observables as form-valued, from which \emph{non--local}  scalar observables are extracted through integration over intermediary  submanifolds. This corresponds indeed to  what is done in Loop Quantum Gravity through the introduction of the Holonomy-Flux algebra. 

The  observables
which commute with the Hamiltonian and with the constraints  correspond to the  \emph{complete  observables} in the sense of  \cite{RovelliPartialobservables,Dittrich} (see also \cite{WestmanSonego}).

\section{ Application to electromagnetism}\label{electromagnetism}

The usual treatment of electromagnetism considers 
 the components $A_\mu$ of the electromagnetic  form $A$ as the dynamical variables, with   the scalar  Lagrangian $L=\undemi F^\munu ~F_\munu$, where  $F_\munu \eqbydef \partial_\mu A_\nu - 
\partial_\nu A_\mu$.
Indices are lowered / raised with the fixed flat Minkowski metric.

1) The usual (non covariant)  analysis proceeds by fixing one time \coord ~$t=x^0$, so that 
$$L=  F^{0i} ~(\partial_0 A_i -  \partial_i A_0)+ 
\undemi F^{ij} ~(\partial_i A_j - 
\partial_j A_i) .$$   We obtain the conjugate momenta
 $P^0\eqbydef \pder{\calL}{\dot {A}_0}=0$  and $P^i=F^{0i}=\partial _0 A_i-\partial _i A_0$.
The first relation appears as the  primary constraint $P^0=0$ and the second inverts as
$\partial _0 A_i=P^i +\partial _i A_0.$
Applying a partial Legendre transform  leads to the Hamiltonian
$$H=\lambda~P^0+\dot {A}_i~P^i-[P^{i} ~P^{i}+ \undemi F^{ij} ~(\partial_i A_j - \partial_j A_i)]$$
$$= \lambda~P^0+( \partial _i A_0)~P^i- \undemi F^{ij} ~(\partial_i A_j - \partial_j A_i).$$

 The primary constraint is second class and generates  the  secondary constraint 
$(P^i)_{,i}=(F^{0i})_{,i}=0$: the Gauss law. Finally,  the motion equations give $\dot{A}_i=(A_0)_i$ and $ \dot{P}^i = - (F{ij})_j$.
This may be synthetized in $F^{\mu\nu}_{,\nu}=0$.

2) The (covariant) multisymplectic analysis starts from the same Lagrangian and,  now,  associates  to each variable $A_\mu$ the four polymomentum components $p^\munu =\undemi (F_\munu-F_{\nu \mu})$. They  obey the constraints $C_\munu=p_\munu+p_{\nu \mu}=0$. The Hamiltonian 
$$\lambda^\munu~ÊC_\munu-\undemi p^\munu~p_\munu,$$   leads to the usual equations, via a multisymplectic analysis (see, \eg \cite{Vey}).

3) Adopting our formalism, we write   $\calL =L~\Vol=\undemi  \d A~\Hodge \d A$ so that  $P= \hodge \d A$, which inverts as $\d A = \hodge  P$:  there is no constraint  and our Hamiltonian takes the form  $\calH =\undemi  P~\hodge P$.

This gives the motion equations 
$$\d  {A} = \pder{\calH}{P}  =\Hodge  {P};$$
$$\d {P}=0,$$ which condense into
$\d \Hodge \d  {A}=0$. 

\section{Application to canonical gravity}\label{GR}
\subsection{Dynamics in the first order formalism}

The dynamical variables are the cotetrad components $e^I$ and the Lorentz connection forms $\omega^{IJ}$, with conjugated polymomenta $P_I$ and $\Pi_{IJ}$. We calculated them  in paper I, 
namely $P_I=0$ and \eql{PIPI}{\Pi_{IJ}=\calP_{IJ}\eqbydef \epsilon_{IJKL}~e^{KL}.}
They generate primary constraints and we  write  the Hamiltonian H-form 
$$\calH=P_I~V^I+(\Pi_{KL} - \calP_{KL}) ~W^{KL}
+P_I~\d e^I+\Pi _{KL}~\d \omega^{KL}- \epsilon_{IJKL}~ e^I~e^J~ ( \d~\omega^{KL}+  (\omega\omega)^{KL}) $$
$$ =P_I~V^I+(\Pi_{KL} - \calP_{KL}) ~W^{KL}
+P_I~\d e^I - \Pi_{KL}      (\omega\omega)^{KL}  $$
with the Lagrange multipliers $V^I$ and $W^{KL}$.

The development of  \eqn{universal} gives the motion equations:
\begin{itemize}
  \item  
  $$\d e^I=\pder{\calH}{P_I}=V^I+\d e^I ,$$
  giving $V^I=0$;
    \item 
\eql{eqomega}{\d \omega^{IJ}=\pder{\calH}{\Pi_{IJ}}=W^{IJ}-(\omega\omega)^{IJ};}
  \item 
\eql{eqP}{\d P_I=-\pder{\calH}{e^I}=2~\epsilon_{KLIJ}~e^J~ W^{KL}.}

The two latter  combine to give 
the secondary constraint\\
 $\d P^I=0=
2~\epsilon_{KLIJ}~e^J~ (\d \omega^{KL}+(\omega\omega)^{KL})$, leading to   zero Ricci curvature;
  \item \eql{eqPi}{\d \Pi _{IJ}=-\pder{\calH}{\omega^{IJ}}=2 ~     \epsilon_{PQK[I}~e^{PQ}~\omega^K_{~~J]}=2~\epsilon_{NKIJ}~e^{NM}~\omega^K_{~~M}} appears as a secondary constraint giving zero torsion, as can be checked using identities   \ref{PIPI} and  \ref{IDID}.

\end{itemize}
  
 \begin{appendices}

 \section{Outline of paper I}\label{outline}
  
 \subsection{Velocity--Histories}
 
Histories
and {velocity-histories} are defined as in the text (\ref{Hamiltonian Histories}).
We  call 
 $  \calS_V \subset \Sect(\VV)$
    the space  of velocity-histories (technically, an \emph{exterior differential system}
\cite{CendraCapriotti}).
Since $\jet$ is canonical,   there is a   one-to-one correspondence between histories and velocity-histories.
 
We  express  the  Lagrangian  dynamics in 
$\calS_V$ rather than in the jet bundle   itself. We treat  $\calS_V$  like an infinite dimensional manifold  where $C$ and $\d C$ play the role of \coord s. 
We  define  
H-maps 
$\calS_V \to \Omega_D$  as generalizations of     functions.  They  form the algebra $\Omega^0(\calS_V)$, and we have  defined  derivations  \wrt their arguments $C$ and $\d C$. We have also defined differential forms on $\calS_V$, forming the spaces $\Omega^r(\calS_V)$, and an exterior   derivative $\D:~\Omega^r(\calS_V)\to \Omega^{r+1}(\calS_V)$, which commutes with $\d$ (occasionally called the \emph{horizontal} exterior derivative).


Dynamics is described through the         \blit{Lagrangian functional}
$$\calL:~\calS_V \to \Omega^n_D:~C_V\eqbydef (C,\d C) \to \calL(C_V),$$  a H-map over $\calS_V$, of type [0,n]. 

We   define  the \blit{historical momentum}  \eql{historical momentum}{P\eqbydef \pder{\calL}{(\d C)}  =P^{\underline{ \mu}}~\Vol_{\underline{ \mu}} } as a [0;n-r-1]-Hmap      admitting the   dual  components $P^{\underline{ \mu}}$. This formula is
written with multi-indexes (see paper I); they reduce to ordinary indices when $C$ is a 0-history; to an antisymmetric pair of indices when $C$ is a 1-history.

Then, applying our differential calculus, we   have 
(wedge products between forms  in $\calD$ are implicitly assumed) 
\eql{momenta}{\D \calL=\D C~
\pder{\calL}{C} +\D Ê(\d   C) ~P   =\D C~(
\dEL{\calL}{C})- \d    \Theta.}
 
We have  defined the EL derivative \eql{ELform}{ \dEL{\calL}{C}\eqbydef   \pder{\calL}{ C } - \epsilon _c~\d \pder{\calL}{(\d C)},}
with $\epsilon _c=(-1)^{\mbox{grade~of~ C}}$; and also 
 the \emph{historical  Lagrange form} (or  Lagrange H-form)  as      the     [1; n-1]-form  
$$\Theta\eqbydef - \D     C  ~P=\D C~\pder{\calL}{(\d C)} .$$  The latter 
gives by derivation the   [2; n-1]-form $ \D \Theta =\D P \wedge \D C$ 
 (implicit wedge product in $\calD$)  which we call the  \emph{symplectic H-form}  (and  $\Theta$   the generalized symplectic potential). This is the \emph{historical version} of the symplectic structure on $\T M$ (see, \eg
\cite{Kharlamov,Cattaneo}). \footnote{ or of the pre-sympletic structure of the evolution space.}

An arbitrary variation of an history  is seen as the result  of the   application of a \vf ~$\delta$ in $\calS_V$ as
$$\delta C =\brak{\D C, \delta};~~\delta ( \d C) =\brak{\D (\d C), \delta}.$$
This leads to 
\eqn{momenta}.
Since the last term in this equation does not contribute to the action, stationarity corresponds to the \ELequ
$\dEL{\calL}{C}=0$.

These equations are explicitely covariant. They    apply   equally well to tD and FT's, and they   include  the case where the  $C$ is a r-history, 
\ie, a   form rather than a  function.

\subsection{Symmetries}

A \vf ~$\delta$ is a symmetry generator  when it does not modifies the action. This  means that it modifies $\calL$ by an exact  form (in $\calD$) $\d X$ only. 
Hence, for a symmetry, $$\delta C~(\dEL{\calL}{C})-\d(\delta C~P)=\d X.$$ Defining the \emph{Noether  current} ([n-1]-H-map)  $j\eqbydef    X + \delta C~P$, we have the conservation law
$$\d j =\delta C~  \dEL{\calL}{C}~\simeq~ 0 \mbox{~(on~shell)}.$$
Locally, $j=\d Q$ which defines the \emph{Noether charge density}  $(n-2)$-H-map $Q$ \cite{Wald}. 


A   diffeomorphism  of $\calD$ is obviously a symmetry  since in that case $\delta\calL =L_\zeta \calL = \d (\innerp{\zeta} \calL)$, where $\zeta$  is the generator.

  The historical Legendre transform (see below) will allow the change of variables $(C,\d C)\leadsto (C,P)$ at the basis of the Hamiltonian formalism.

\subsection{Legendre transform }

 The    (usual) Legendre transform       transports the dynamics from $\VV$   to \YY. It is   defined as  the    fiber-preserving map \cite{Gotay 1991}
$$T_L:~\VV \to \YY:~(x^\mu,\varphi,v_\mu)\leadsto (x^\mu,\varphi,p^\mu, \pi),$$ 
here for a scalar field, in adapted \coord s.   
\footnote{ It admits  a   \emph{restricted} version
$$\VV \leadsto \widetilde{\YY}:~ (x^\mu,\varphi,v_\mu)\leadsto (x^\mu,\varphi,p^\mu ).$$ }
It may be   non     invertible,
what  is expressed by primary  constraints. We assume now a non degenerate Legendre transform,    constraints are discussed   in the examples.  

We lift the Legendre transform  
  to    the \emph{historical Legendre map} which applies to sections, the duality 
$$\calT_L:~\calS_V  \to  \calY  :~C=(C,\d C) \leadsto Y=(C,P)  $$  between 
velocity-histories and   Hamiltonian histories.       This results from  the simple remark that a fiber-preserving map between \fb s induces a map between their spaces of sections.        
Concretely, the velocity history $C_V$ is transformed,   by composition with  $T_L$,      as 
$Y= T_L \circ C_V$.

\subsection{The  historical Hamiltonian}

We  define    the \blit{historical  Hamiltonian} on the historical phase space 
\eql{Hform}{\calH:~\calY\to \Omega^n(\calD):~ Y \to \calH(Y)  =      \Lambda ^i~\Gamma_i +\Pi~\Vol +P ~  \d C -\calL} 
(wedge product assumed).   
 In this expression,  $\d C$ and $\calL$ are expressed as functionals of $C$ 
and~$P$, as  far as allowed by inversion of the Legendre map, so that  $\calH$ is   a
  [0;n]--Hmap.
  The $ \Lambda ^i$ and $\Gamma_i$ are   Lagrange multipliers and constraints,  which are now   defined as Hmaps also (see  illustrations in examples).
This definition holds for r--histories.

 \section{Details of Calculations}\label{Details}
\begin{itemize}
  \item 
For a \emph{scalar field}, a field-history is a zero-form $C$ on $\calD$.
The momentum  is a 3 form $P=P^\mu ~Ê\Vol_\mu$ (its Hodge dual $\Hodge P=P^\mu ~\d x^\mu$ is a 1-form).

The Hamiltonian functional  is a [0;4]-H-map $\calH=h~\Vol$.
Its external derivative $$\D \calH =
\pder{\calH}{C}~\D C
+\pder{\calH}{P}~\D P
...=
\pder{h}{C}~\D C~\Vol
+\pder{h}{P^\mu}~\D P^\mu~\Vol ... ,$$
so that 
$$\pder{\calH}{C}=\pder{h}{C} ~\Vol,~~~~~~ \pder{\calH}{P}  = \pder{h}{P^\mu}~\d x^\mu.$$

  \item 

For a \emph{one-form  field}, a field--history is a one form $C=C_\alpha~\d x^\alpha$ on $\calD$.
The momentum  is a 2 form $P=P^{\alpha\mu} ~Ê\Vol_{\alpha\mu}$. 

The Hamiltonian is a [0;4]-H-map $\calH=h~\Vol$.
Its external derivative $$\D \calH =
\pder{\calH}{C}~\D C
+\pder{\calH}{P}~\D P
...=
\pder{h}{C_\alpha}~\D C_\alpha~\Vol
+\pder{h}{P^{\alpha\mu}}~\D P^{\alpha\mu}~\Vol ... ,$$
so that 
$$\pder{\calH}{C}= \pder{h}{C_\alpha}~\Vol_\alpha~,
,~~~~~~ \pder{\calH}{P}  = \pder{h}{P^{\alpha\mu}}~\d x^\mu~\d x^\alpha.$$

\end{itemize}

 \section{Multi-index notations} \label{multiindex}

For a r-history we write 
$$C=   
C_{\underline{\mu}}~\d  ^{\underline{\mu}},$$ where
$\underline{\mu}$ means the (antisymmetrized) sequence $(\mu_1,...,\mu_r)$ and 
$\d  ^{\underline{\mu}} $ means 
$\d x^{ \mu_1}...\d x^{ \mu_r}.$

Similarly,  the momentum,  
$$P=
P^{\underline{\nu}}~\Vol _{\underline{\nu}}=
P^{\underline{\nu}}~\epsilon  _{\underline{\nu}, \underline{\rho}}~\d^{\underline{\rho}};~~~~{\underline{\nu}} ~\eqbydef \nu_1,...,\nu_{r+1},$$ with  
$\epsilon  _{\underline{\nu}, \underline{\rho}}~\eqbydef \epsilon  _{\nu_1,...,\nu_{r+1}, \rho_1,...,\rho  _{n-r-1}};~~~~~
\Vol _{\underline{\nu}} ~\eqbydef \innerp {\partial _{\underline{\nu}}}\Vol= \epsilon  _{\underline{\nu}, \underline{\rho}}~\d ^{\underline{\rho}}$;\\ involving 
the multivector $\partial_{\underline{\nu}}=(\partial_{\nu_1}, ..., \partial_{\nu_{r+1}})$.

We expand similarly    a [0, R]-Hmap as
$$F=F _{\underline{\alpha}}~\d  ^{\underline{\alpha}}.$$  

It results, \eg, 
    \eql{dddC}{
\pder{F}{ C}=
\pder{F_{\underline{\alpha}}}{ 
C_{\underline{\mu} }}~
  (\innerp{ \partial_{\underline{\mu}}}
 \d ^ {\underline{\alpha}})  ,
}  
   \eql{dddP}{
   \pder{F}{ P}=
    \epsilon^{\underline{\nu}\underline{\rho}}~
~   \pder
   {F_{\underline{ \alpha}}}{ P^{\underline{\nu}}}~
 (    \innerp{\partial_{\underline{\rho}}}
\d ^{ \underline{ \alpha}}). 
      }

Note that the validity of these formulas implies conditions for   the grades, namely
$R\ge r$ and $R\ge n-r-1$ respectively.
We will restrict  to such  situations sufficient for our purpose, although  generalizations are possible. 

 
 \section{An identity}\label{Idd}
To prove the    identity   :
\eql{IDID}{\epsilon_{JKAB}~e^{JI}~\omega^K_{~~I}=- \epsilon_{J[A~MN}~e^{MN}~\omega^J_{~~B]},}
  we use $$ \Hodge e^{IJ}  = \undemi \epsilon ^{IJ}_{~~~MN}~e^{MN};~~
\undemi e^{IJ}  =- \epsilon ^{IJ}_{~~~MN}~(\Hodge e^{MN}).$$
Then,
$$\epsilon_{JKAB}~e^{IJ}~\omega^K_{~~I}= \undemi \epsilon_{JKAB}~\epsilon ^{IJ}_{~~MN}~(\Hodge e^{MN})~\omega^K_{~~I}=
\eta_{BC}~( \Hodge e^{KC})~\omega _{KA}-  \eta_{AC}~(\Hodge e^{KC})~\omega _{KB} $$
$$= -  \epsilon  _{K[BMN}~e^{MN}~\omega^K_ {~A]}, ~QED $$

Similarly, \eql{IdD}{\epsilon_{IJKL}~e^{IJ}~\omega^K_{~~A}~\omega^{AL}= -  \epsilon_{JKMN}~e^{IJ}~\omega^{K}_{~~I}~\omega^{MN}.}

\end{appendices}

 \end{document}